\documentstyle[aps,pre,multicol,epsf]{revtex}

\begin{document}
\date{\today}
\title{Thermodynamic Cross-Effects from Dynamical Systems}
\author{L\'aszl\'o M\'aty\'as,$^{(1)}$
Tam\'as T\'el,$^{(1)}$ and
J\"urgen Vollmer$^{(2),}$\cite{present} }
\address{
(1) Institute for Theoretical Physics,
E\"otv\"os University,
P. O. Box 32, 
H-1518 Budapest,
Hungary. 
\\
(2) Center for Nonlinear Phenomena and Complex Systems,
Universite Libre de Bruxelles,
C.P. 231 --- Campus Plaine,
B-1050 Brussels,
Belgium.
}
\maketitle
\begin{abstract}
We give a thermodynamically consistent description of simultaneous heat and 
particle transport, as well as of the associated cross-effects, in the 
framework of a chaotic dynamical system, a generalized multibaker map. 
Besides the density, a second field with 
appropriate source terms is included in order to mimic,
after coarse graining,  a spatial 
temperature distribution and its time evolution. 
A new expression is derived for the irreversible entropy
production in a steady state, as the average of
the growth rate of the relative density, a unique
combination of the two fields.
\end{abstract}
\draft

\widetext 
\begin{multicols}{2}
%
The relation between transport processes and chaotic models with 
only a few degrees of freedom (cf.~\cite{Dorf} for recent 
reviews) became a subject of active research since the rapid 
progress in dynamical-system theory started in the early eighties. 
First, it was shown that such models can faithfully describe 
particle transport, and become compatible on the macroscopic level 
with appropriate macroscopic transport  equations \cite{Grossmann}. 
Later it was found that the irreversible behavior of these 
processes, expressed, e.g., by their entropy production 
\cite{CELS,Gent,VTB97,VTB98} or fluctuation 
relations \cite{ECM,RTV}, can properly be 
obtained in a more restricted class of models only --- in 
particular, if one wants to keep them low dimensional. 
Multibaker maps 
\cite{G,TVB,Gent,VTB97,VTB98,GD98,TG98,MTV99}, 
the extensions of baker maps \cite{Ott} to 
a macroscopically long array of mutually connected unit cells  
turned out to be an effective tool to understand
the origin of irreversibility on the level of dynamical systems. 
Besides the possibility of explicit calculations, they lead 
to general findings \cite{Gent,VTB98,RTV} valid also 
outside the realm of multibakers. 

When describing transport in the presence of an external field and/or a
density gradient, consistency with the thermodynamic entropy balance
could only be obtained for multibaker maps with a time-reversible, local 
dissipation mechanism \cite{VTB97,VTB98} 
(a brief discussion of this notion will be given below). 
This requirement was interpreted as 
mimicking a thermostatting algorithm (a Gaussian thermostat, 
cf.~\cite{books}), which is widely applied in nonequilibrium 
molecular dynamics (NEMD). 
The entropy balance was found to hold for a coarse-grained entropy 
based on a density averaged over regions of small spatial extension. 
A recent paper by Tasaki and Gaspard \cite{TG98} shows 
that analogous results can be obtained for area-preserving multibaker 
maps with an energy-dependent phase-space volume. 
This energy, however, is strictly connected to the potential of an 
external field, and not considered as an independent 
driving force. 

\begin{figure}
{\hfill
\epsfbox{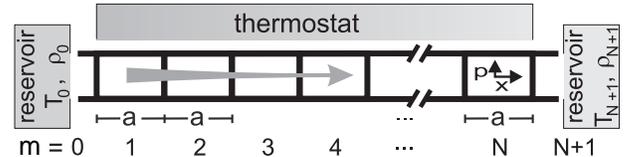} 
\hfill} 
\caption[]{\narrowtext
Graphical illustration of the transport process considered.
The system is attached to reservoirs inducing particle and heat 
currents as indicated by the arrow, and along its extent heat can 
be exchanged with a thermostat. 
\label{fig-cartoon}}
\end{figure}

In the present paper, our aim is to study transport generated by 
two independent driving forces: density and temperature gradients. 
In addition, we allow for a constant external field. 
We are intending to describe a quasi one-dimensional system of 
finite length attached at the two ends to different reservoirs, and
possibly in thermal contact with a 
thermostat along its extension (cf. Fig.~\ref{fig-cartoon}). 
In this general setting we show that thermoelectric phenomena, 
i.e., cross effects due to the simultaneous presence of two 
independent driving forces, and the entropy balance can properly 
be modelled by an elementary dynamical system. 

The multibaker map describes transport along the $x$ direction. 
It represents a deterministic dynamics [the $(x,p)$ dynamics] 
in the {\em  single-particle phase space\/} 
of a weakly interacting many-particle system. 
The cell size $a$ partitions 
the $x$ axis into regions which are sufficiently large 
to characterize the state inside such a 
cell by thermodynamic state variables and small enough to neglect 
variation of these variables on the length scale of the cells 
({\em local-equilibrium approximation\/}). 
Thus, $a$ plays the role of a minimum allowed macroscopic resolution. 
The state of the many-particle system is represented
by the (particle) density $\varrho$ and a so called ``kinetic-energy'' 
density $\varrho T$, whose average over cells is related to a local
temperature.  
For multibaker maps the kinetic-energy density is considered as an
independent field, i.e., our discussion does not rely on the apparance
of a momentum conjugated to the $x$ variable. 
The $(x,p)$ dynamics drives the time evolution of the fields, leading
to a mesoscopic description of the transport process.  
A possible dependence of this dynamics on local thermodynamic
averages, and the presence of a source term of kinetic energy
introduces a coupling of the fields. 
  
\begin{figure}
\vspace*{1.5mm}                      
\epsfbox{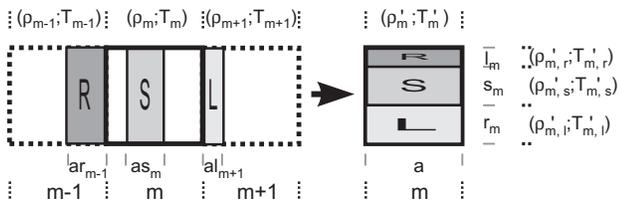} 
\caption[]{\narrowtext
The action of the multibaker map on 
the coordinates $(x,p)$ over a time unit $\tau$. 
The values of $\varrho(x,p)$ and ${T(x,p)}$ on the cells and strips 
[cf. Eqs. (\ref{eq:rho_update}) and (\ref{eq:T_update})] are given 
on the margins for the initial conditions discussed in the text. 
\label{fig-mbaker}}
\end{figure}

The detailed definition of the model is as follows: 
The multibaker map acts on a domain in the $(x,p)$ plane consisting  
of $N$ identical cells labelled by $m$ (Fig.~\ref{fig-cartoon}). 
Here, $x$ is a position variable, and $p$ is a momentum-like variable 
needed to set up a reversible deterministic dynamics. 
Every cell has a width $a$ and height $b \equiv 1$. 
After every time unit $\tau$, every cell is divided into three 
columns (Fig.~\ref{fig-mbaker}) with respective widths 
   $a l_m$, $a s_m$ and $a r_m$ 
fulfilling    
   $l_m + s_m + r_m = 1$.
The right (left) column of width
     $a r_{m}$ 
    ($a l_{m}$) 
is uniformly squeezed and stretched into a strip of width $a$ and of height
     ${l_{m+1}}$ 
    (${r_{m-1}}$) 
in the right (left) neighbouring cell.
The middle one preserves its area.
The map is time reversible in the sense that the Jacobian 
$l_{m+1}/r_m$ for a motion from cell $m$ to $m+1$ is reciprocal 
to that of the motion from cell $m+1$ to $m$.  
The $(x,p)$ dynamics is volume preserving when an initial condition is
recovered, but it is contracting on average (since motion in the
direction of the external field is connected with contraction), and
hence it is dissipative.  
Except for the coupling to reservoirs at the ends the map is 
one-to-one on its domain. 
It drives the density $\varrho(x,p)$ 
and the kinetic-energy density $\varrho (x,p) T(x,p)$. 
Both are advected, but --- in order to be able to capture a
local heating of the system --- the density $\varrho T$ is also 
multiplied by a factor 
$(1+\tau q_m)$ with $q_m$ depending only on the averages 
in cell $m$ and in its neighbors. 
The dynamics of both densities is governed by the Frobenius-Perron 
equation \cite{Dorf} of the $(x,p)$ map, but after each iteration 
the $\varrho T$ values in cell $m$ are multiplied by 
$(1+\tau q_m)$. 
The fields $\varrho$ and $T$ evolve into fractal 
distributions whose asymptotic forms 
are described by {\it different\/} invariant measures.  
In general, the width $l_m, s_m, r_m$ of the columns may depend on the 
average values of the fields in the cell and its neighbors so that the  
widths may vary in time and space.

In the spirit of nonequilibrium thermodynamics, we also 
consider the coarse-grained fields $\varrho_m$ and $T_m$ interpreted 
as the density and the local temperature of cell $m$, respectively.
$\varrho_m$ and $\varrho_m T_m$ are obtained as averages
of $\varrho(x,p)$ and of the kinetic-energy density $\varrho(x,p) T(x,p)$
over cell $m$.

We now discuss the time evolution of $\varrho$ and $T$. 
For the explicit calculation we start with a density $\varrho(x,p)$
and a specific kinetic energy $T(x,p)$ taking constant values
$\varrho_m$ and $T_m$ in each cell $m$.  
This is convenient from a technical point of view, and does not 
lead to a principal restriction of the domain of validity of the 
model as discussed in \cite{VTB98,GD98}. 
After one step of iteration, the fields will be piecewise constant 
on the strips defined in Fig.~\ref{fig-mbaker}. 
Due to continuity, the density takes the respective values 
\begin{equation}
   \varrho_{m,r}' =  \frac{r_{m-1}}{{l_{m}}} \; \varrho_{m-1} ,
\;\;\;\;
   \varrho_{m,s}' = \varrho_{m} ,
\;\;\;\;
   \varrho_{m,l}' =  \frac{l_{m+1}}{{r_{m}}} \; \varrho_{m+1}.  
\label{eq:rho_update}
\end{equation}
(The prime always indicates quantities evaluated after one time step.) 
Besides the advection by the $(x,p)$ dynamics, 
the updated values for $T$ on the strips contain the source term: 
\begin{eqnarray}
   T_{m,r}' = T_{m-1} \; \left[ 1 + \tau q_m \right] ,
\nonumber \\ 
   T_{m,s}' = T_{m}   \; \left[ 1 + \tau q_m \right] ,
\label{eq:T_update}
\\ 
   T_{m,l}' = T_{m+1} \; \left[ 1 + \tau q_m \right] .
\nonumber 
\end{eqnarray}

In the $x$ variable, 
the cell-to-cell dynamics of the model is equivalent to a random
walk with fixed step length $a$ and local transition
probabilities $r_m$ and $l_m$ over time unit $\tau$.
Such random walks are characterized by the drift $v_m$ and the 
diffusion coefficient $D$, which stay finite in the macroscopic 
limit $a,\tau \rightarrow 0$. 
To be consistent with a diffusion type equation for the density,
the transition probabilities have to scale \cite{TVB} as 
\begin{eqnarray}
   r_{m} - l_{m}  = (\tau/a)\, v_m  ,
\qquad 
   r_{m} + l_{m} = (2\tau/a^{2})\,D .
\label{eq:rm+-lm}
\end{eqnarray}

In order to account for the effect of temperature gradients on 
the local $(x,p)$ dynamics, we allow in the present paper
for a location dependence of the drift $v_m$, while 
the diffusion coefficient is  kept spatially homogeneous. 
The location dependence of the drift is due to a dependence on the cell 
temperature $T_m$ and on its discrete gradient: 
$
v_m=v_m \left(  T_m, (T_{m+1}-T_m)/a  \right) $.

The entropy $S_m$ of cell $m$ is related to the cell density 
$\varrho_m$. 
We use the common information-theoretical form 
\begin{equation}
   S_m 
= 
   -a \varrho_m 
   \ln\left[ \varrho_m / \varrho^{\star}(T_m) \right] . 
\label{Sm}
\end{equation}
It is measured in units of Boltzmann's constant, and involves 
a temperature dependent reference density 
$
   \varrho^{\star}(T)
\equiv 
  {{T}^{\gamma}} 
$, 
in close analogy to an ideal gas. 
Here $\gamma$ is a free parameter.
In Ref.~\cite{MTV99} it will be demonstrated that 
only this choice of the reference density can be consistent with 
thermodynamics.
In the present paper, we take the form of (\ref{Sm}) for granted and
explore in how far the model leads to proper macroscopic expressions
for quantities appearing in the local entropy balance, i.e., for 
(i) the irreversible entropy production, 
(ii) the current, 
(iii) the heat and entropy currents, and 
(iv) the transport coefficients. 

Next, we work out the
evolution equations of the coarse-grained fields and of the  
entropy $S_m$. 
Due to (\ref{eq:rho_update}), the change of the cell 
density in a time unit $\tau$ becomes 
\begin{equation}
   ({\varrho_m}' - \varrho_m) / {\tau} 
=
- ({j_m - j_{m-1}})/{a} 
,
\label{eq:rhoprime}
\end{equation}
where 
\(   
   j_m = (a/\tau) \; ( r_m \varrho_m - l_{m+1} \varrho_{m+1} ) 
\)   
denotes the discrete current density. 

Similarly, an update of the average temperature in the cells  
can be calculated based on the fact that $\varrho_m T_m$ is a 
kinetic-energy density. 
Consequently
[cf.~Eqs.~(\ref{eq:rho_update},\ref{eq:T_update})], 
the change $\varrho_m ' T_m ' - \varrho_m T_m$ 
per time step is obtained as 
\begin{equation}
   \frac{\varrho_m^{'} T_m^{'} - \varrho_m T_m}{\tau} 
= 
    \varrho_m^{'} T_m^{'} \frac{q_m}{1 + \tau q_m} 
 -  \frac{j_m^{(\varrho T)} - j_{m-1}^{(\varrho T)}}{a} , 
\label{eq:wmprime1}
\end{equation} 
where 
\(   
   j_m^{(\varrho T)} 
= 
   T_m j_m  -  \varrho_{m+1}\, ({T_{m+1}-T_m})/{a} \; 
   ({a^2}/{\tau})\, l_{m+1} 
\)   
is a discrete heat current, and 
   $\varrho_m^{'} T_m^{'} q_m/(1 + \tau q_m)$ 
represents a source of kinetic energy arising from irreversible 
heating.

In order to find an evolution equation for the 
entropy, we write $S'_m - S_m$
in the form of a discrete entropy-balance equation 
(cf.~\cite{VTB98}) 
\begin{equation}
   S'_m - S_m = \Delta_e S_m + \Delta_i S_m , 
\label{eq:micro-balance}
\end{equation}
where 
\begin{mathletters}
\begin{eqnarray}
   \Delta_e S_m & \equiv & {S_m^{(G)}}' - S_m^{(G)} ,
\label{eq:DeltaeSm}
\\
   \Delta_i S_m & \equiv & 
      ({S_m}' - {S_m^{(G)}}') -  ({S_m} - {S_m^{(G)}}) , 
\label{eq:DeltaiSm}
\end{eqnarray}
\end{mathletters}
correspond to the entropy flux and the irreversible entropy
production, respectively.
Following Refs.~\cite{VTB98} and in harmony with (4), we define 
the Gibbs entropy as 
\begin{equation}
   S^{(G)}
=
   -\int \hbox{d} x \,\hbox{d} p \; \varrho(x,p) \;
      \ln\left[\varrho(x,p) \; T(x,p)^{-\gamma} \right] .
\label{eq:SG}
\end{equation}
In view of (8a), the entropy flux is the temporal change of 
the Gibbs entropy. 
Moreover, (\ref{eq:DeltaiSm}) is a meaningful measure for the rate 
of  irreversible entropy production, since according to the 
information theoretic interpretation of entropies, it describes the 
increase per unit time of the lack of information due to 
coarse-graining. 
Hence it is positive. 
The dependence of $S_m - S_m^{(G)}$ on
details of the coarse graining drops out when taking the 
time derivative \cite{VTB98}. 

In view of these considerations, we find 
the following expressions for the entropy flux and the 
rate of irreversible entropy production [cf.\ Fig.~\ref{fig-mbaker} and
Eqs.~(\ref{eq:rho_update},\ref{eq:T_update})]  
\end{multicols} 
\widetext 
\vspace*{-9mm}
\begin{flushleft} {\rule{88mm}{0.2mm}\rule{0.2mm}{2mm}} \end{flushleft}
\begin{mathletters} 
\begin{eqnarray} 
   \frac{\Delta_e S_m}{\tau} 
& = &  - \frac{a}{\tau} 
\left[ 
   (\varrho_m^{'} - \varrho_m) \; 
      \ln\left( {\varrho_m}  
      T_m^{-\gamma} \right) 
+
   \varrho_m^{'} \ln \frac{T_{m,s}'^{-\gamma}}{T_m^{-\gamma}}
+
   \varrho_{m-1} r_{m-1} 
      \ln\left( \frac{\varrho_{m,r}'}{\varrho_m} 
                \frac{T_{m,r}'^{-\gamma}}{T_{m,s}'^{-\gamma}} 
         \right) 
+ 
   \varrho_{m+1} l_{m+1} 
      \ln\left( \frac{\varrho_{m,l}'}{\varrho_m}
                \frac{T_{m,l}'^{-\gamma}}{T_{m,s}'^{-\gamma}} 
         \right) 
\right], 
\label{eq:SGdot} 
\\ 
   \frac{\Delta_i S_m}{\tau} 
& = & 
 \frac{a}{\tau} \; \bigg[  
   - \varrho_m' \ln\left( 
   \frac{\varrho_m' T_m'^{-\gamma}}{\varrho_m T_{m,s}'^{-\gamma}} 
              \right)
   + \varrho_{m-1} r_{m-1} 
      \ln\left( 
         \frac{\varrho_{m,r}' T_{m,r}'^{-\gamma}}{\varrho_m T_{m,s}'^{-\gamma}}
      \right) 
   + \varrho_{m+1} l_{m+1} 
      \ln\left( 
         \frac{\varrho_{m,l}'T_{m,l}'^{-\gamma}}{\varrho_m T_{m,s}'^{-\gamma}}
         \right)    
\bigg] 
. 
\label{eq:DiSm} 
\end{eqnarray} 
\label{eq:Delta-ei-S}
\end{mathletters} 
\vspace*{-7mm}
\begin{flushright}{\rule[-2mm]{0.2mm}{2mm}\rule{88mm}{0.2mm}}\end{flushright}
\vspace*{-7mm}
\begin{multicols}{2} 
In the last equation we made use of the fact that due to the 
originally homogeneous field distributions, the initial 
coarse-grained and microscopic densities coincide such that $S_m = S_m^{(G)}$. 

A detailed discussion of the entropy current will be 
given elsewhere \cite{MTV99}.  
Concerning the entropy production
in a steady state we only mention the following interesting relation:  
The irreversible change of the specific entropy per unit time 
appears to be the average of the growth rate $\sigma_{\varrho/\varrho^*}$
of the {\em relative density\/} $\varrho(x,p)/\varrho^*(x,p)$, i.e., 
$ \Delta_i S_m/(a \tau \varrho_m) 
  = \langle \sigma_{\varrho/\varrho^*} \rangle_{ss}$, 
where the growth rate is defined as 
\begin{equation} 
   \sigma_{\varrho/\varrho^*} (x,p)  \equiv  \frac{1}{\tau} \,
      \ln{\frac{[\varrho(x,p)/\varrho^*(x,p)]_\tau}
               {[\varrho(x,p)/\varrho^*(x,p)]_0}} . 
\label{sigmarhoT}
\end{equation}
This rule is a direct extension of the result for uniform temperature
\cite{VTB98}, where only the growth rate of the density appeared. 

Next, we evaluate the entropy balance (for a general 
time-dependent state) in the macroscopic limit. 
It is obtained by refining the partition of the finite total 
length $L=N a$ of the chain. 
Technically this means 
\(
   a,\tau \rightarrow 0
\)
with fixed $L$ and diffusion coefficient $D$.
[By this condition a macroscopic limit $v(T, \partial_x T)$ is obtained 
for the total drift $v_m$ ]. 
The macroscopic spatial coordinate is defined as 
   $x \equiv a m$, and  
the spatial 
dependence of any  field $\psi=\varrho$ or $T$ can 
be expressed as 
\begin{equation}
    \psi_{m\pm 1}
\rightarrow
   \psi \pm a \partial_x \psi  +  \frac{a^2}{2} \partial_x^{2} \psi .
\label{eq:n+-1}
\end{equation}

By dividing Eq.~(\ref{eq:micro-balance}) by $a \tau$, 
we find in this limit a balance equation for the 
entropy density $S_m/a \rightarrow s$ in the form 
\(
   \partial_t s = \Phi + \sigma^{\rm (irr)} 
\)
\begin{mathletters}
with
\begin{eqnarray}
   \Phi
& \equiv & \lim_{a,\tau \rightarrow 0} \frac{\Delta_e S_m}{a\, \tau} 
= 
   - \partial_x j^{(s)}
   + \Phi^{\rm (thermostat)}.
\label{eq:Phi-therm}
\\ 
   \sigma^{\rm (irr)} 
& \equiv & 
 \lim_{a,\tau \rightarrow 0} \frac{\Delta_i S_m}{a\, \tau} 
= 
   \frac{j^2}{\varrho D} 
+  \gamma \varrho D  \left[ \frac{\partial_x T}{T} \right]^2, 
\label{eq:sirr-therm}
\end{eqnarray}
\label{eq:sirr-Phi-therm}
\end{mathletters}
Here, $\Phi^{\rm (thermostat)}$ stands for the entropy flux into 
the thermostat, and $j$ and $j^{(s)}$ are the current and 
entropy current densities, respectively. 
The former is obtained from the time evolution (\ref{eq:rhoprime}) of 
the density, which becomes an advection-diffusion equation in the 
thermodynamic limit with a current 
\(
   j  =   \varrho \,  v  \left(  T, \partial_x T  \right)
            -  D  \,  \partial_x \varrho . 
\)
The entropy current can be expressed as 
\begin{equation}
   j^{(s)}  = -
   \left[ 1 + \ln\left(\varrho T^{- \gamma} \right) \right] j 
 -
   \gamma \varrho D\,  \partial_x T / T , 
\label{eq:therm-currents}
\end{equation}
and for the flux into the thermostat we obtain 
\begin{equation}
   \Phi^{\rm (thermostat)} 
= 
   \gamma \varrho   q - v j / D.
\end{equation} 

The expression for the currents and the irreversible entropy 
production are in harmony with thermodynamics \cite{dGM}. 
A comparison shows that 
$  
 \lambda \equiv  \gamma \varrho D 
$  
plays the role of the heat conductivity, 
and 
$  
   -T \left(1 + \ln\left[\varrho T^{-\gamma}) \right] \right)/e 
$  
is the Peltier coefficient ($e$ stands for the unit charge).
The appropriate form of 
$
v\left(  T, \partial_x T  \right)
$ 
follows from Onsager's reciprocity 
relation \cite{MTV99}. 
Thus, we have expressed all the kinetic coefficients with system 
parameters. 
   
For a system isolated from the thermostat (a non-thermostatted system) one has 
   $\Phi^{\rm (thermostat)} = 0$, 
which fixes $q$ to be $v j/\lambda$. 
This choice of $q$ exactly corresponds
to classical thermodynamics, where the full entropy flux can 
be written as the negative divergence of the entropy current. 
In this case, the heat leads to a rise of temperature, reflected 
in the source term 
   $ qT = vjT /\lambda$ 
of the evolution equation 
$\partial_t T = q T + \partial_x (\lambda\, \partial_x T)/(\gamma \varrho) -
(j/\varrho)\, \partial_x T$ 
for the local temperature [i.e., the macroscopic limit of
(\ref{eq:wmprime1})]. 
The case $q=0$ corresponds to a Gaussian thermostat 
commonly applied in NEMD simulations \cite{books}. 
Other choices of $q$ describe stationary states stabilized by 
appropriate entropy fluxes to (or from) the 
thermostat.
Temperature profiles,  which are not steady in thermodynamics, appear then
as steady states by a proper choice of $q$, implying
$\Phi^{\rm (thermostat)} \neq 0$,  which represents a  
generalization of the classical thermodynamic entropy balance.

To conclude, we point out that to our knowledge transport with 
cross effects and the associated entropy balance has never been 
treated in the framework of dynamical systems where both the 
properties of the underlying dynamics and of the thermodynamic time evolution 
can explicitly be worked out 
(for early efforts based on Gaussian thermostats, cf.~\cite{Evans}). 
It is remarkable that a thermodynamic description of 
transport driven by two independent forces 
could be derived from a model as simple as a 
piecewise-linear map. 
The minimum requirements on the model in order to 
be consistent with thermodynamics are found to be: 
{(i)} a time-reversible dissipation mechanism, 
{(ii)} inclusion of a new field $T$ describing a local kinetic-energy density, 
{(iii)} defining the entropy with a $T$-dependent reference density, 
{(iv)} inclusion of a source term in the time evolution of the field
$T$, which can model the effect of Joule's heat. 

Observations which seem to be valid beyond the frame of 
multibaker models are: 
(a) Dissipation and thermostatting  play different roles. 
With time-reversible dissipation we can also describe systems 
isolated from a thermostat 
(b) Different choices of the source term in the temperature 
dynamics correspond to changes in the coupling of the system to 
the thermostat. 
(c) A proper expression for the total irreversible entropy 
production should involve the steady-state average 
of the relative density $\varrho/\varrho^*$, instead of 
phase-space contraction. 
(d) In macroscopic steady states, when the coarse-grained densities do
not change in time, the transition probabilities and the source term
in the dynamics are fixed to time-independent values. 
In this case, one finds a {\em stationary\/} entropy balance based on
an inhomogeneous low-dimensional mapping acting on a spatially
extended (i.e., macroscopic) domain. 



We are grateful to G.\ Nicolis for many enlightening 
discussions, and to our referees for helpful comments. 
Support from the Hungarian Science Foundation (OTKA T17493, 
T19483),the German-Hungarian D125 Cooperation, and the TMR-network 
{\em Spatially extended dynamics\/} is acknowledged. 

\vspace*{-5mm}


\end{multicols}

\end{document}